\pdfoutput=1
\documentclass[journal=langmuir,manuscript=article,amsmath]{achemso}
\usepackage[version=3]{mhchem} 

\author{L. G. L\'opez}
\author{A. J. Ramirez-Pastor}
\email{lglopez@unsl.edu.ar, antorami@unsl.edu.ar} \affiliation[UNSL] {Departamento de
F\'{\i}sica, Instituto de F\'{\i}sica Aplicada, Universidad
Nacional de San Luis-CONICET, Ej\'ercito de los Andes 950,
D5700BWS San Luis, Argentina}

\title{Adsorption of Self-Assembled Rigid Rods on Two-Dimensional Lattices}

\begin{document}

\begin{abstract}
Monte Carlo (MC) simulations have been carried out to study the
adsorption on square and triangular lattices of particles with two
bonding sites that, by decreasing temperature or increasing
density, polymerize reversibly into chains with a discrete number
of allowed directions and, at the same time, undergo a continuous
isotropic-nematic (IN) transition. The process has been monitored
by following the behavior of the adsorption isotherms (chemical
potential $\mu$ as a function of the surface coverage $\theta$)
for different values of lateral interaction
energy/temperature. The numerical data were compared with
mean-field analytical predictions and exact functions for
noninteracting and 1D systems. The obtained
results revealed the existence of three adsorption regimes in
temperature. (1) At high temperatures, above the critical one
characterizing the IN transition at full coverage $T_c(\theta=1)$,
the particles are distributed at random on the surface and the
adlayer behaves as a noninteracting 2D system. (2)
At very low temperatures, the asymmetric monomers adsorb forming
chains over almost the entire range of coverage, and the adsorption
process behaves as a 1D problem. (3) In the intermediate regime,
the system exhibits a mixed regime and the filling of the lattice
proceeds according to two different processes. In the first stage,
the monomers adsorb isotropically on the lattice until the IN
transition occurs in the system and, from this point, particles
adsorb forming chains so that the adlayer behaves as a 1D fluid.
The two adsorption processes are present in the adsorption
isotherms, and a marked singularity can be observed that separates both
regimes. Thus, the adsorption isotherms appear as sensitive
quantities with respect to the IN phase transition, allowing us (i) to
reproduce the phase diagram of the system for square lattices and
(ii) to obtain an accurate determination of the phase diagram for
triangular lattices.

\end{abstract}

\newpage

\section{1. Introduction}

The adsorption of gases on solid surfaces is a topic of
fundamental interest for various applications
\cite{Steele,Keller,Toth}. From the theoretical point of view, the
process can be described in terms of the lattice-gas theory
\cite{Steele,Keller,Toth,Dash,Dash1,Shina,BinLan,Patry} or its
generalizations, including lateral interactions between adsorbed
particles, surface heterogeneity, multilayer adsorption, etc.
\cite{Rudzi,Jaro,Rudzi1,Hill,Zhdanov,Tovbin,SS4,Adamson,Gregg}. In
all cases, the symmetry of the interactions is preserved and,
consequently, isotropic adlayers are observed. However, most real
components, from proteins to ions \cite{DeYoreo} to the wide
variety of recently synthesized nanoparticles
\cite{Glotzer,Glotzer1}, interact via anisotropic or ``patchy"
attractions and clear signals of symmetry breaking have been
observed. Experimental realization of such systems is growing. An
example of real patchy particles is presented in Ref.
\cite{Cho}. Such particles offer the possibility to be
used as building blocks of specifically designed self-assembled
structures \cite{Whitesides,Glotzer,Glotzer1,Zhang1,Zhang2}.

The term ``molecular self-assembly" may be used to refer to
spontaneous formation of an ordered molecular overlayer on a
surface. Molecular self-assembly on surfaces \emph{via} weak but
selective noncovalent interactions, offers a promising bottom-up
approach to fabricate highly organized systems from instructed
molecular building blocks. In this way, the engineering of
supramolecular arrays, with desired functionalities, on metal
surfaces can be performed \cite{Barth}. If, in addition, such
systems are capable of undergoing phase transitions, then their
functional properties may be increased.

Recently, several research groups reported on the assembly of
colloidal particles in linear chains. Selectively functionalizing
the ends of hydrophilic nanorods with hydrophobic polymers, Nie et
al. reported the observation of rings, bundles, chains, and
bundled chains \cite{Nie}. In another experimental study carried
out by Chang et al. \cite{Chang}, gold nanorods were assembled
into linear chains using a biomolecular recognition system.

Another prominent case is the self-assembly of flexible
one-dimensional coordination polymers on metal surfaces, i.e. in
Ref. \cite{Heim}, where the researchers employed a
\emph{de novo} synthesized porphyrin module to construct
one-dimensional (1D) Cu-coordinated polymers on Cu(111) and
Ag(111) surfaces.

In a direct relation with the present work, Clair et al.
\cite{Clair} investigated the self-assembly of terephthalic acid
(TPA) molecules on the Au(111) surface. Using scanning tunneling
microscopy, the authors showed that the TPA molecules arrange in
one-dimensional chains with a discrete number of orientations
relative to the substrate. However, in the experimental studies of
chains self-assembled on a surface with a discrete number of
orientations, orientational ordering transitions were not studied.

In this context, the main objective of this paper is to investigate the
adsorption process in a system composed of monomers with two attractive
(sticky) poles that self-assemble reversibly into polydisperse chains and,
at the same time, undergo an orientational transition.
In a recent series of papers \cite{Tavares,PRE4,Tavares1,JCP13,JCP14,Almarza,PRE9,Almarza1}, the critical behavior of this system
has been widely studied.  Except mean-field calculations \cite{JCP14}, these
studies showed the existence of a continuous phase transition along the
entire coexistence curve. However, the universality class of the model has
been a subject of controversy \cite{PRE9,Almarza1}. It was shown that the system under
study represents an interesting case where the use of different statistical
ensembles (canonical or grand canonical) leads to different and
well-established universality classes ($q=1$ Potts-type or $q=2$ Potts-type,
respectively) \cite{PRE9}. The present work goes a step further addressing the study of the
adsorption isotherms in different regions of the phase diagram,
emphasizing their behavior at critical and subcritical
temperatures. The adsorption isotherm appears as a sensitive
quantity to the phase transition, allowing a very accurate
determination of the phase diagram.  These findings may be of
interest for theoretical and experimental studies on critical
adsorption phenomena.

It is clear that the model considered here is highly idealized and
is not meant to reproduce a particular experimentally studied
system. However, the understanding of simple models with
increasing complexity might be a help and a guide for future
experimental investigations. This work represents an effort in
that direction.

The paper is organized as follows. In Sec. 2, lattice-gas model
and theoretical formalism (mean-field approximation and exact
functions for non-interacting and 1D systems) are presented. Section 3 is devoted to
describe the Monte Carlo simulation scheme. The
analysis of the results and discussion are given in Sec. 4.
Finally, the conclusions are drawn in Sec. 5.

\section{2. Lattice-Gas Model and Theory}
\label{Lattice-Gas Model}

As in Refs.
\cite{Tavares,PRE4,Tavares1,JCP13,JCP14,Almarza,PRE9,Almarza1}, a
system of self-assembled rods with a discrete number of
orientations in two dimensions is considered. The substrate is
represented by a square or triangular lattice of $M = L \times L$
adsorption sites, with periodic boundary conditions. $N$ particles
are adsorbed on the substrate with $m$ possible orientations along
the principal axis of the array, being $m=2$ for square lattices
and $m=3$ for triangular lattices (see Fig. 1). These particles
interact with nearest-neighbors (NN) through anisotropic
attractive interactions. Thus, a cluster or uninterrupted sequence
of bonded particles is a self-assembled rod. Then, the grand
canonical Hamiltonian of the system is given by
\begin{equation}
H = w \sum_{\langle i,j \rangle} |\vec{r}_{ij} \cdot
\vec{\sigma}_i||\vec{r}_{ji} \cdot \vec{\sigma}_j| \ {\rm div} \ 1
- \left(\mu- \epsilon_o \right) \sum_i |\vec{\sigma}_i|,
\label{Ham}
\end{equation}
where $\langle i,j \rangle$ indicates a sum over NN sites; $w$
represents the NN lateral interaction between two neighboring $i$
and $j$, here $w<0$ and the energy is lowered by an amount $|w|$
only if the NN monomers are aligned with each other and with the
intermolecular vector $\vec{r}_{ij}$; $\vec{\sigma}_j$ is the
occupation vector, with $\vec{\sigma}_j=0$ if the site $j$ is
empty and $\vec{\sigma}_j= \hat{x_i}$ if the site $j$ is occupied
by a particle with orientation along the $x_i$-axis; and
$\epsilon_o$ is the adsorption energy of an adparticle on a site.
In the present work, $\epsilon_o$ was set equal to zero (without
any loss of generality) and the chemical potential
$\mu$ is the only parameter that determines the strength of the
adsorption. The operation ${\rm div}$ 1, which is redundant in the case of
square lattices, (i) avoids additional lateral interactions
\cite{foot0} that promote the condensation of the monomers in the
triangular lattice; and (ii) restricts  the attractive couplings only to
those pairs of NN monomers whose orientations are aligned with
each other and with the monomer-monomer lattice direction, in line
with the model on the square lattice. The sign of $\mu$ is chosen
so that negative $\mu$ makes vacancies favorable. At
$\mu=+\infty$, vacancies are suppressed and we wind up with the
full lattice case. At $\mu=-\infty$, all we have are vacancies and
the ground state is just vacant.

The introduction of intermolecular forces brings about the
possibility of phase transitions \cite{Hill,Goldenfeld,Yeomans}.
Among the common types of phase transitions are, condensation of
gases, melting of solids, transitions from paramagnet to
ferromagnet and order-disorder transitions. From a theoretical
point of view, when NN interactions are present, an extra term in
the partition function for interaction energy is required. With
this extra term, only partition functions for the whole system can
be written. Ising \cite{Ising} gave an exact solution to the
one-dimensional lattice problem in 1925. All other cases are
expressed in terms of series solution \cite{Hill,Domb}, except for
the special case of two-dimensional lattices at half-coverage,
which was exactly solved by Onsager \cite{Onsager} in 1944. Close
approximate solutions in dimensions higher than one can be
obtained, and the most important of these is the Bragg-Williams
approximation (BWA). BWA is the simplest mean-field treatment for
interacting adsorbed particles, even in the case of asymmetric
particles as studied here. In this theoretical framework, the
isotherm equation takes the form \cite{Hill,Lang4}
\begin{equation}
K(c) e^{(\mu-\epsilon_o-c w \theta)/k_BT} =
\frac{\theta}{1-\theta}, \label{isomf}
\end{equation}
where $T$ is the temperature, $k_B$ is the Boltzmann constant,
$\theta=N/M$ is the surface coverage and $K(c)$ represents the
number of available configurations (per lattice site) for a
monomer at zero coverage. The term $K(c)$ is, in general, a
function of the connectivity of the lattice $c$ and the structure
of the adsorbate ($c=$ 4 and 6 for square and triangular lattices,
respectively). It is easy demonstrate that $K(c)=2$ for square
lattices and $K(c)=3$ for triangular lattices \cite{IECR1}.

In the limit of high temperatures ($w/k_BT \rightarrow 0$),
particles adsorb randomly on the surface and Eq. (\ref{isomf})
reduces to the well-known Langmuir isotherm \cite{Hill}:
\begin{equation}
K(c)e^{(\mu- \epsilon_o)/k_BT} = \frac{\theta}{1-\theta}.
 \label{isolang}
\end{equation}

On the other hand, the phase diagram reported in Ref.
\cite{JCP14} indicates that, by decreasing temperature or
increasing density, the monomers polymerize reversibly and form
linear chains on the surface. In this limit, it is more convenient
to describe the system from the exact form of the adsorption
isotherm corresponding a interacting monomers on 1D lattices
\cite{Hill}:
\begin{equation}
K(c)e^{(\mu- \epsilon_o-w)/k_BT} =
\frac{b-1+2\theta}{b+1-2\theta},
 \label{iso1D}
\end{equation}
where $b=\{1-4\theta \left(1-\theta \right) \left[ 1-
\exp{(-w/k_BT)}  \right] \}^{1/2}$ and $K(c)=1$ for 1D lattices.

\section{3. Monte Carlo Simulation}

The thermodynamic properties of the present model were
investigated following a standard importance sampling MC method in
the grand canonical ensemble \cite{BINDER}. The procedure is as
follows. For a given pair of values of $T$ and $\mu$, an initial
configuration with $N$ monomers adsorbed at random positions and
orientations (on $M$ sites) is generated. Then an
adsorption-desorption process is started, where the lattice sites
are tested to change its occupancy state with probability given by
the Metropolis rule \cite{METROPOLIS}: $P = \min
\left\{1,\exp\left( - \beta \Delta H \right) \right\}$, where
$\Delta H$ is the difference between the Hamiltonians of the final
and initial states and $\beta=1/k_BT$. Insertion and removal of
monomers, with a given orientation, are attempted with equal
probability. For this purpose, an axis orientation (with
probability $1/2$ for the square lattice, and $1/3$ for the
triangular lattice) and a lattice site are chosen at random. If
the selected lattice site is unoccupied, an attempt to place a
monomer (with the orientation previously chosen) on the site is
made. If, instead, the site is occupied, then the algorithm checks
the orientational state of the adsorbed monomer and if this
coincides with the previously chosen orientation, an attempt to
desorb the particle is performed; otherwise, the trial ends. A
Monte Carlo Step (MCS) is achieved when $M$ sites have been tested
to change its occupancy state. The equilibrium state can be well
reproduced after discarding the first $s_0 = 10^5-10^6$ MCS. Then,
averages are taken over $s = 10^5-10^6$ successive configurations.

In this framework, total and partial adsorption isotherms can be
obtained as simple averages:
\begin{equation}
\theta = \frac{1}{M} \sum_i^M \langle |\vec{\sigma}_i| \rangle,
\end{equation}
and
\begin{equation}
\theta_{x_i} = \frac{1}{M} \sum_i^M \langle |\vec{\sigma}_i \cdot
\hat{x_i} | \ {\rm div} \ 1 \rangle,
\end{equation}
where $\theta$ represents the total surface coverage;
$\theta_{x_i}$ denotes the partial surface coverage referring to
the adparticles with orientation along the $x_i$-axis
($\theta=\sum_{i=1}^{c/2} \theta_{x_i}$); and $\langle \dots
\rangle$  means the time average over the Monte Carlo simulation
runs.

All calculations were carried out using the parallel cluster BACO
of Universidad Nacional de San Luis, Argentina. This facility,
located at Instituto de Física Aplicada, Universidad Nacional de
San Luis-CONICET, San Luis, Argentina, consists of 50 CPUs each
with an Intel Core i7 processor running at 2.93 GHz and 512MB of
RAM per core.

\section{4. Results and Discussion}

In the present section, the main characteristics of the
thermodynamic functions given in \ref{isomf,iso1D},
will be analyzed in comparison with simulation results for a
lattice-gas of asymmetric monomers on 2D lattices. The
computational simulations have been developed for square and
triangular $L \times L$ lattices, with $L$ = 100, and periodic
boundary conditions. With this lattice size we verified that
finite-size effects are negligible.

In order to understand the basic phenomenology, it is instructive
to begin by discussing the behavior of the simulation adsorption
isotherms for square lattices and different values of the
magnitude of the lateral interactions [Fig. 2]. Curves from right
to left correspond to $w/k_BT=$0, -1, -2, -3, -4, -5, -6, -7, -8,
-9 and -10, respectively. As it can be observed, isotherms shift
to lower values of the chemical potential and their slope
increases as the ratio $|w|/k_BT$ increases. This behavior is
typical of attractive systems. However, a notable difference is
observed with respect to the case of symmetric monomers
\cite{Hill}, where two nearest-neighbors particles interact with
an interaction energy $w$. In fact, a clear discontinuity (jump)
is observed in the adsorption isotherms corresponding to symmetric
monomers for interactions above a critical value $w>w_c$ (in
absolute values). The inset of Fig. 2 shows a scheme of the
described process. The marked jump, which has been observed
experimentally in numerous systems, is indicative of the existence
of a first-order phase transition. In this situation, the only
phase which one expects is a lattice-gas phase at low coverage,
separated by a two-phase coexistence region from a
``lattice-fluid" phase at higher coverage \cite{foot1}.

On the other hand, the surface coverage varies continuously with
the chemical potential for all cases shown in Fig. 2, and there is
no evidence of the existence of a first-order phase transition. In
other words, the mechanism for which the asymmetric monomers
polymerize reversibly into chains at low temperatures (high values
of $|w|/k_BT$) is not associated with the existence of a
first-order phase transition in the adlayer. This point will be
discussed in the next.

In Ref. \cite{JCP14}, the phase diagram of the system  was
calculated by using Monte Carlo simulations on a square lattice.
The results are reproduced in Fig. 3(a) to facilitate the reader
with a self-contained presentation. Circles joined by lines
represent Monte Carlo simulation data obtained by L\'opez et al.
\cite{JCP14}. This line is a line of critical transitions that
separates regions of isotropic and nematic stability.

In the following, we address the study of the adsorption isotherms
in different regions of the phase diagram [different paths
indicated in Fig. 3(a)], emphasizing their behavior at critical
and subcritical temperatures.

Fig. 4(a) reports a comparison between theory (lines) and Monte
Carlo simulation (symbols) for particles adsorbed on square
(squares) and triangular (triangles) lattices. Two
well-differentiated limits have been considered in the figure: (i)
non-interacting particles (full symbols): $w/k_BT=0$, path not
shown in Fig. 3(a); and (ii) strongly interacting particles (open
symbols): $w/k_BT=-8$, path (1) in Fig. 3(a), and $w/k_BT=-10$,
path (2) in Fig. 3(a).

In the first case [(i)], lines correspond to results from \ref{isolang} with $K(c)=2$ (square lattices) and $K(c)=3$
(triangular lattices). As expected, asymmetric monomers adsorb at
random on the surface and an excellent agreement is obtained
between theory and Monte Carlo simulation. The same behavior has
been observed for values of $k_BT/|w|$ above
$k_BT_c(\theta=1)/|w|$, being $T_c(\theta=1)$ the critical
temperature characterizing the IN transition at full coverage. A
typical configuration of the adlayer in this regime [$\theta=0.5$
and $k_BT/|w|=0.65$, point $I$ in Fig. 3(a)] is shown in Fig. 3(b).

In the second case [(ii)], lines represent data from \ref{iso1D}. From a first inspection of the curves it is
observed that (a) the results corresponding to square and
triangular lattices coincide in a unique curve; and (b) simulation
data agree very well with theoretical results obtained for 1D
lattices. These findings contribute to the understanding of the
polymerization transition. Namely, in the case of strongly
interacting particles (very low temperature limit), the IN phase
transition occurs at very low coverage [see points $A$ and $B$ in
Fig. 3(a)] and the asymmetric monomers adsorb forming chains over
almost all range of coverage. A typical configuration of the
adlayer in this regime [$\theta=0.6$ and $k_BT/|w|=0.2$, point $J$
in Fig. 3(a)] is shown in Fig. 3(c).

Then, (1) the adsorption process behaves as a 1D problem, (2) the
shape of the triangular and square isotherms are
indistinguishable, and (3) at variance with the behavior observed
for symmetric monomers (see inset of Fig. 2), the system studied
here does not show a first-order phase transition at low
temperature (it is well-known that no phase transition develops in
a 1D lattice-gas \cite{Hill}).

In Ref. \cite{JCP14}, the phase diagram corresponding to
square lattices was also obtained by mean-field calculations. The
results showed the existence of a coexistence region between a
low-coverage isotropic phase and a high-coverage nematic one at
low temperatures. The presence of this coexistence region
(first-order phase transition) was completely at variance with the
observed numerical simulation results. The study in Fig. 4(a)
allows now to understand the mean-field prediction. Namely, the
system goes to a 1D adsorption as the temperature decreases and,
as it is well-known, mean-field approximation incorrectly predicts
a phase transition in one dimension and low temperatures
\cite{Hill}. The comparison in Fig. 4(b) confirms this argument.
In the figure, simulation isotherms (symbols) obtained for square
lattices and three different values of $w/k_BT$ (as indicated) are
compared with the corresponding ones obtained from \ref{isomf} (lines). The mean-field curves lead to the
characteristic van der Waals loops and, consequently, to the
prediction of a first-order phase transition \cite{Hill}.

Summarizing the discussion above, the adsorption process passes
from a 2D problem (at high temperatures) to a 1D one (at very low
temperatures). At the intermediate regime, the system behaves in a
mixed regime and the simulation isotherms are not well fitted,
neither by \ref{isolang} nor by \ref{isomf} (data not
shown here). These findings provide very important information
about the critical behavior of the system and, as it will be seen
in the following, suggest a method to obtain the phase diagram of
the system through adsorption measurements.

Let us suppose now that the system is found in the point $F$ of
the phase diagram (intermediate temperature, isotropic region);
with increasing coverage (at constant $k_BT/|w|$, in this case
$k_BT/|w|=1/3$), the system passes (through the critical point
$G$) from the non-oriented to the nematic phase at point $H$. The
path ($F-G-H$) is shown on the corresponding adsorption isotherm
in Fig. 5(a). The figure also show a similar study for
$k_BT/|w|=1/4$, path ($C-D-E$) in Fig. 3(a). A detailed inspection
of the curves in Fig. 5(a) reveals the existence of singularities
in the critical points $D$ and $G$. To confirm this, the
adsorption isotherms were differentiated with respect to the
chemical potential. As an example, the results for the case
$k_BT/|w|=1/3$ are shown in the inset of Fig. 5(b), where the
critical coverage can be determined from the position of the
maximum. The same methodology has been extended to other
temperatures and the values obtained for the critical coverage
(solid stars) agree very well with previous results from Ref.
\cite{JCP14}.

The existence of a singularity in the adsorption isotherms
indicates that a dramatic change in the adsorption process takes
place in the system. One way to visualize this situation is to
consider the partial adsorption isotherms. As an example, total
and partial adsorption isotherms are plotted in Fig. 5(b) for
square lattices and $k_BT/|w|=1/3$. $\theta_{x_1}(\mu)$ and
$\theta_{x_2}(\mu)$ represent the fraction of particles adsorbed
along the $x_1$-axis (horizontal axis) and the fraction of
particles adsorbed along the $x_2$-axis (vertical axis),
respectively. Two well-differentiated adsorption regimes can be
observed: (i) for $\theta<\theta_c$, particles are adsorbed
isotropically on the surface and
$\theta_{x_1}(\mu)=\theta_{x_2}(\mu)$, and (ii) for
$\theta>\theta_c$, particles are adsorbed forming chains along one
of the lattice axes (vertical axis in the case of the figure).

The procedure illustrated in Fig. 5 was used to obtain the
temperature-coverage phase diagram corresponding to a triangular
geometry. The results are shown in Fig. 6. Full circles represent
data obtained from the inflection points in the adsorption
isotherms. The figure also includes recent MC data by Almarza et
al. \cite{Almarza} (open circles). The rest of the curve
separating isotropic and nematic stability (dashed line) was built
on the basis of the previous results on the behavior of the system
at low temperatures (see Fig. 4). Namely, the critical properties
corresponding to square and triangular lattices coincide in the
low-temperature (coverage) regime. The complete phase diagram of
triangular lattices has been reported here for the first time. A
more exact determination of $T_c$ based on Monte Carlo simulations
and finite-size scaling theory is in progress.

For comparative purposes, Fig. 6 includes the critical line
corresponding to square lattices (solid line). Even though the
shapes of the curves are similar, the critical temperature
corresponding to a given density ($\theta > 0.1$) is higher for
square lattices than for triangular lattices.

In summary, the existence of singularities in the adsorption
isotherms, which are induced by the phase transition, provides a
way to determine the temperature-coverage phase diagram from
adsorption experiments. These findings may be instructive for
future theoretical and experimental studies on adsorption.

\section{5. Conclusions}

In the present paper, the main adsorption properties of
self-assembled rigid rods on square and triangular lattices have
been addressed. The results were obtained by using Monte Carlo
simulations, mean-field theory and exact calculations in one
dimension. According to the present analysis, the behavior of the
system is characterized by the following properties:

\begin{itemize}

\item[1)] At high temperatures, above the critical one
characterizing the IN transition at full coverage $T_c(\theta=1)$,
the particles are distributed isotropically on the surface, the
adlayer behaves as a 2D system and an excellent agreement is
observed between Monte Carlo simulation and theoretical results
from Langmuir isotherm.

\item[2)] At very low temperatures, asymmetric monomers adsorb
forming chains over almost all range of coverage and the
adsorption process passes from a 2D problem to a 1D one. Then, the
results corresponding to square and triangular lattices coincide
in a unique curve, simulation data agree very well with exact
theoretical results obtained for 1D lattices, and the system
studied here does not show a first-order phase transition at low
temperatures (it is well-known that no phase transition develops
in a 1D lattice-gas \cite{Hill}).

\item[3)] Mean-field calculations \cite{JCP14} showed the
existence of a coexistence region (first-order phase transition)
between a low-coverage isotropic phase and a high-coverage nematic
one, at complete variance with the observed numerical simulation
data. The result described in the item above allows now to
understand the mean-field prediction. Namely, the system goes to a
1D adsorption as the temperature decreases and, as it is
well-known, mean-field approximation incorrectly predicts a phase
transition in one dimension \cite{Hill}.

\item[4)] At intermediate temperatures, the system exhibits a
mixed behavior, the 2D and 1D adsorption processes are present in
the adsorption isotherms and a marked singularity can be observed
separating both regimes. The measurement of the point at which
this singularity occurs (pronounced maximum in the derivative of
$\theta$ with respect to $\mu$) allows an accurate determination
of the critical coverage characterizing the IN phase transition.

\item[5)] The simple analysis described in the last item
reproduces the previously obtained phase diagram for square
lattices \cite{JCP14} and provides a good approximation for the
case of triangular lattices. A more detailed and accurate study,
including the variation of critical exponents and the
determination of the phase diagram through finite-size scaling
analysis, is being undertaken for triangular lattices.

\end{itemize}

\acknowledgement

This work was supported in part by CONICET
(Argentina) under project number PIP 112-200801-01332; Universidad
Nacional de San Luis (Argentina) under project 322000 and the
National Agency of Scientific and Technological Promotion
(Argentina) under project PICT-2010-1466.

\newpage

\section*{Figure captions}

\noindent Fig. 1. Schematic representations of isotropic (left) and nematic
(right) configurations of self-assembled rigid rods on square
(top) and triangular (bottom) lattices. Here, the lattice size is
L=30.
\\[12pt]

\noindent Fig. 2. Simulation adsorption isotherms for square lattices and
different values of the magnitude of the lateral interactions.
Curves from right to left correspond to $w/k_BT=$0, -1, -2, -3,
-4, -5, -6, -7, -8, -9 and -10, respectively. Inset: Scheme
showing the behavior of the adsorption isotherms for symmetric
monomers and different values of $w/k_BT$ (above and below the
critical value $w_c /k_BT$).
\\[12pt]

\noindent Fig. 3. (a) Phase diagram of the system on a square lattice:
circles joined by lines represent Monte Carlo simulation data
obtained by L\'opez et al. \cite{JCP14} and stars correspond to
results obtained from the singularities in the adsorption
isotherms. The meaning of the dotted lines and the points $A-J$ is
discussed in the text. (b) Simulation snapshot obtained in the
isotropic region [point $I$ in Fig. 3(a)]. (c) Same as (b) for the
nematic phase [point $J$ in Fig. 3(a)]. In both cases, the lattice
size is $L = 100$ and vertical and horizontal rods are shown in
different colors for clarity.
\\[12pt]

\noindent Fig. 4. (a) Comparison between theoretical (lines) and simulation
(symbols) isotherms for particles adsorbed on square ($c=4$) and
triangular ($c=6$) lattices. Curves correspond to different values
of $w/k_BT$ as indicated. In the case of $w/k_BT=0$ [$w/k_BT=-8$
and -10], lines correspond to results from \ref{isolang}
[\ref{iso1D}]. (b) Lattice coverage $\theta$ versus relative
chemical potential $\mu/k_BT$ for square lattices and different
values of $w/k_BT$ as indicated. Solid lines correspond to MFA
[\ref{isomf}] and symbols represent results from Monte Carlo
simulation.
\\[12pt]

\noindent Fig. 5. (a) Simulation adsorption isotherms obtained for
$w/k_BT=-3$ and $w/k_BT=-4$ as indicated. Points $C$, $D$, $E$,
$F$, $G$ and $H$, corresponding to the phase diagram in Fig. 3(a),
are indicated on the curves. (b) Total and partial adsorption
isotherms obtained for $w/k_BT=-3$. Inset: Derivative of the total
isotherm with respect to the chemical potential. The critical
coverage can be determined from the position of the
maximum.
\\[12pt]

\noindent Fig. 6. Phase diagram corresponding to asymmetric particles
adsorbed on square (solid line) and triangular (dashed line)
lattices. Full circles represent results obtained from the
singularities in the adsorption isotherms and open circles
correspond to MC data by Almarza et al. \cite{Almarza}.
\\[12pt]

\newpage

\begin{figure}[t]
\includegraphics[width=16cm,clip=true]{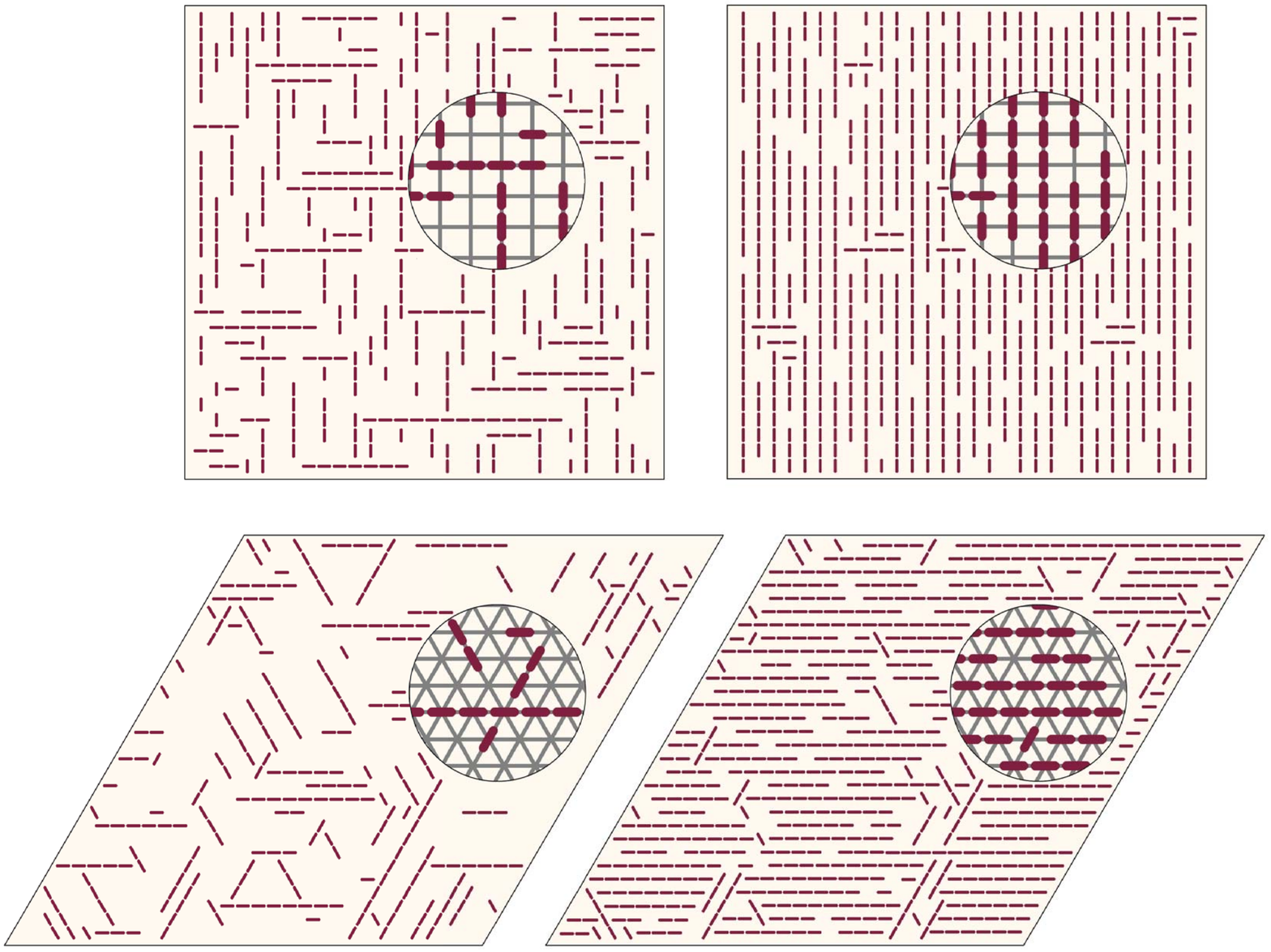}
\caption{}\label{figure1}
\end{figure}

\newpage

\begin{figure}[t]
\includegraphics[width=16cm,clip=true]{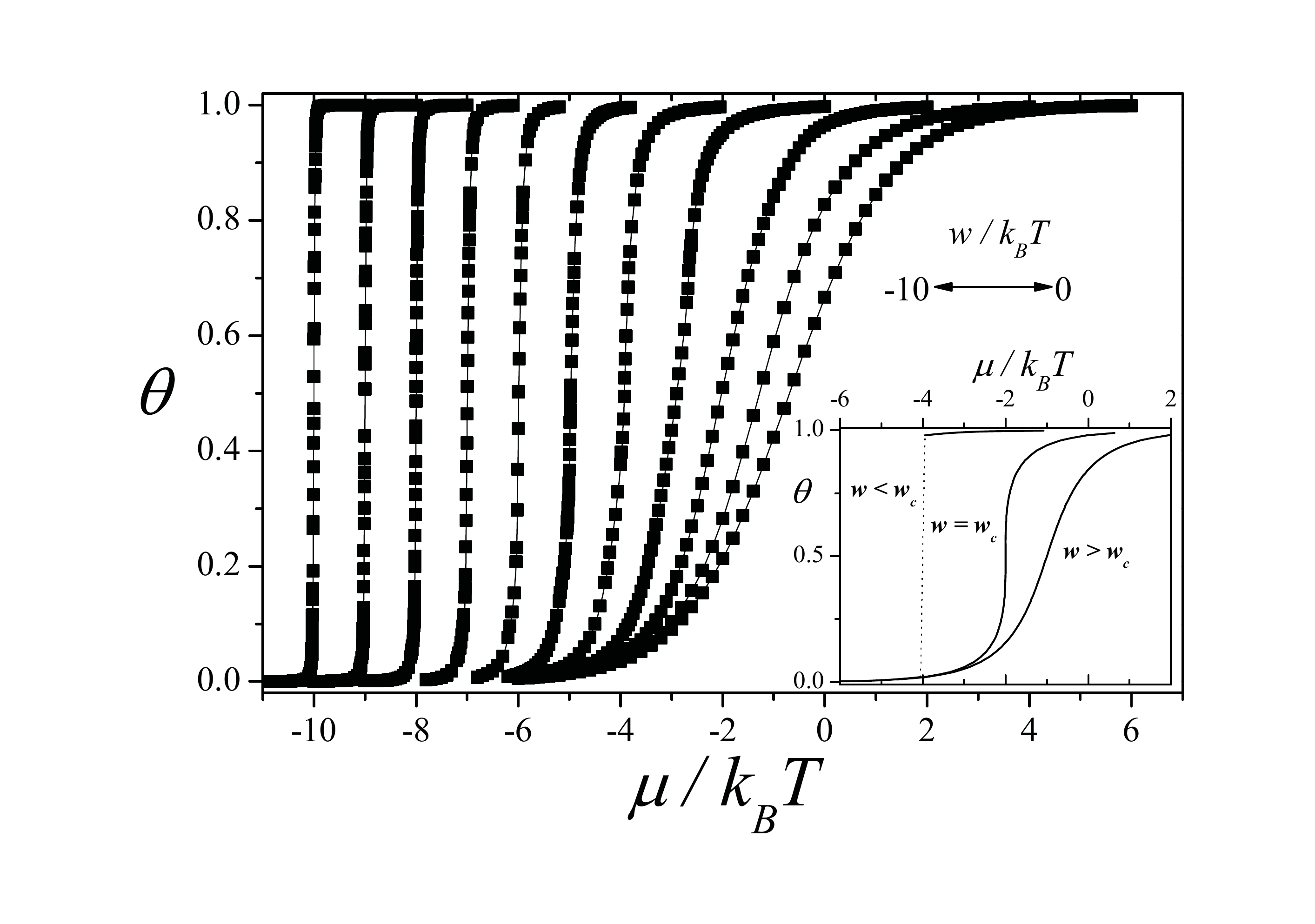}
\caption{}\label{figure2}
\end{figure}

\newpage

\begin{figure*}[t]
\includegraphics[width=16cm,clip=true]{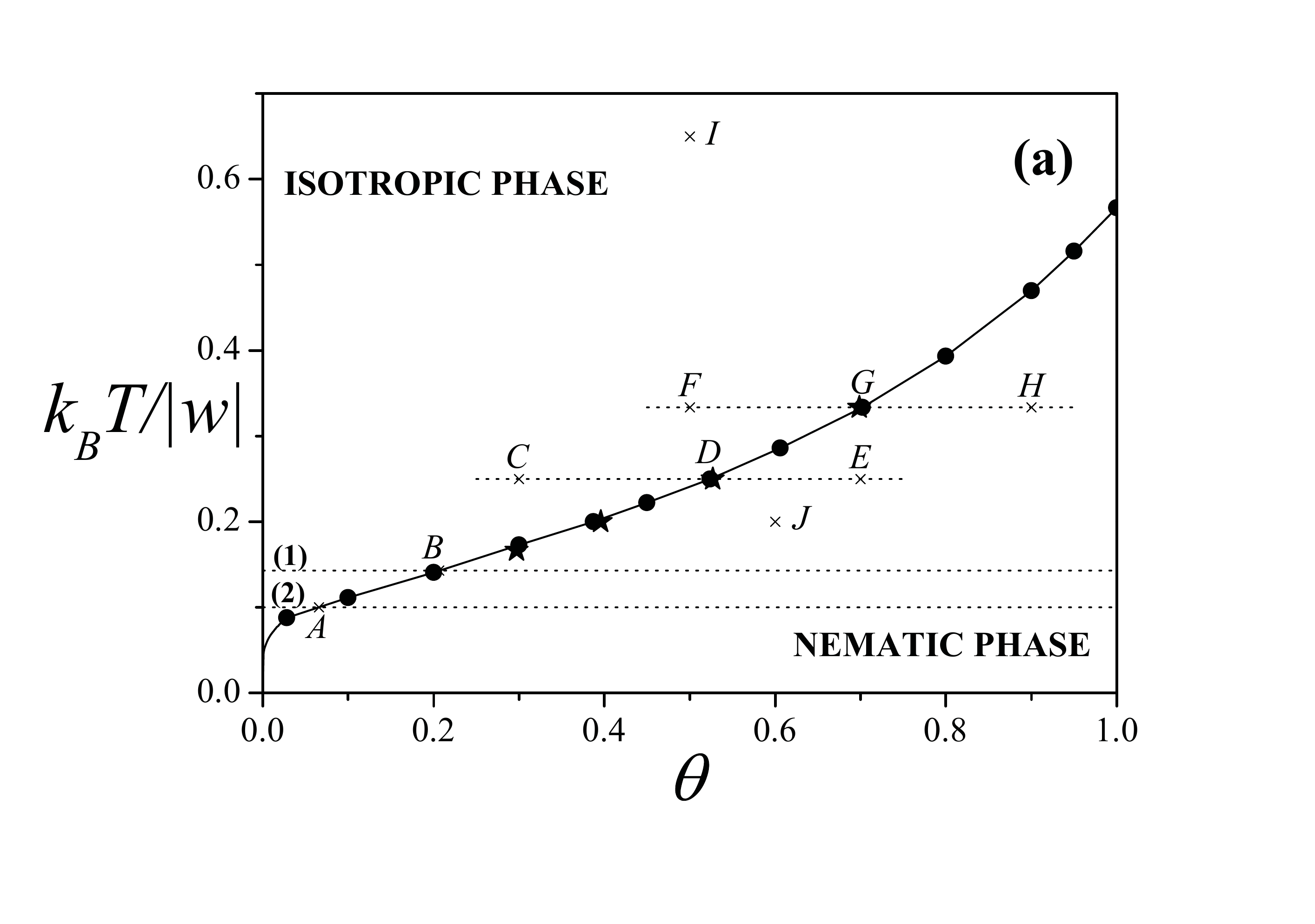}
\includegraphics[width=8cm,clip=true]{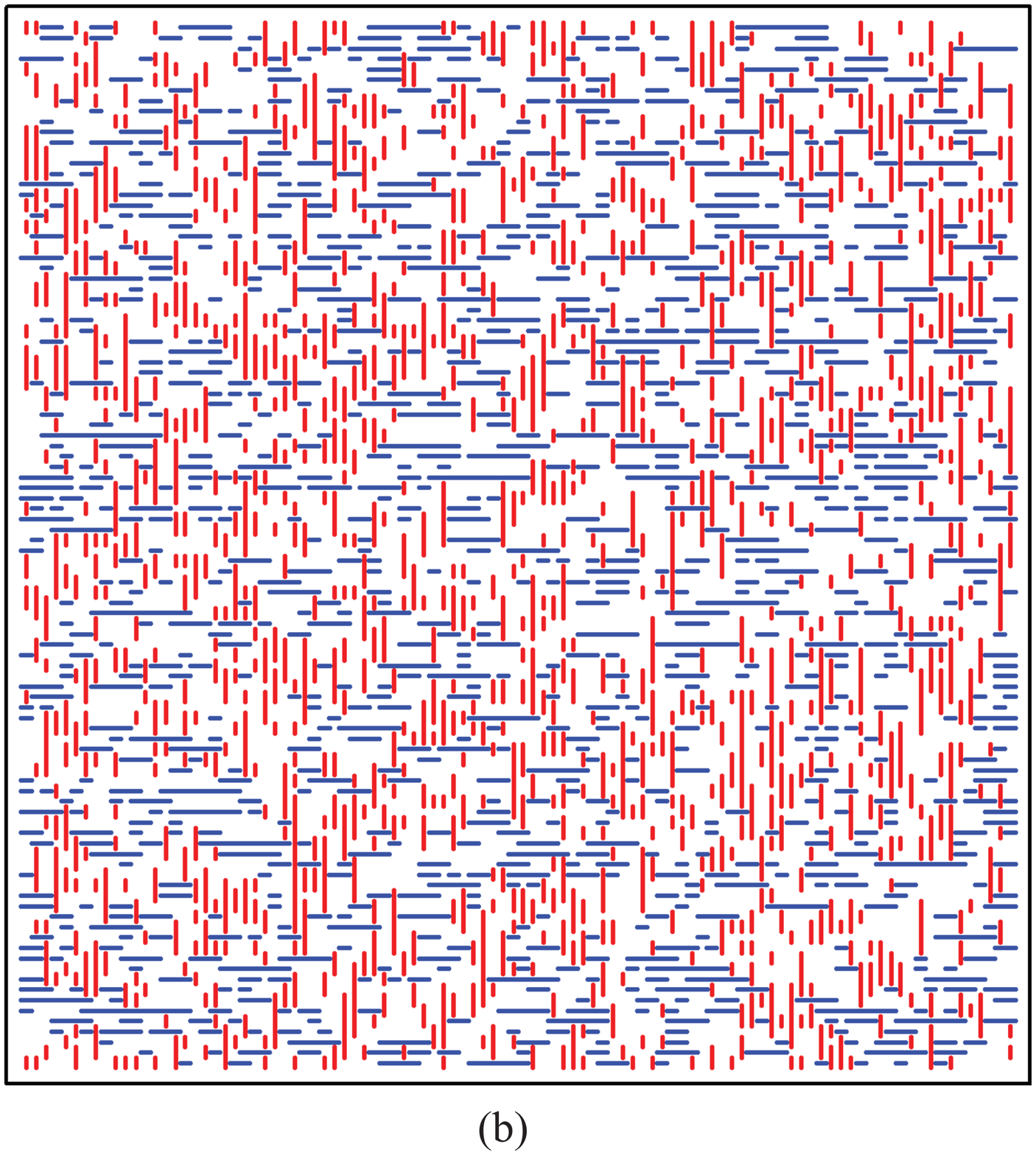}
\includegraphics[width=8cm,clip=true]{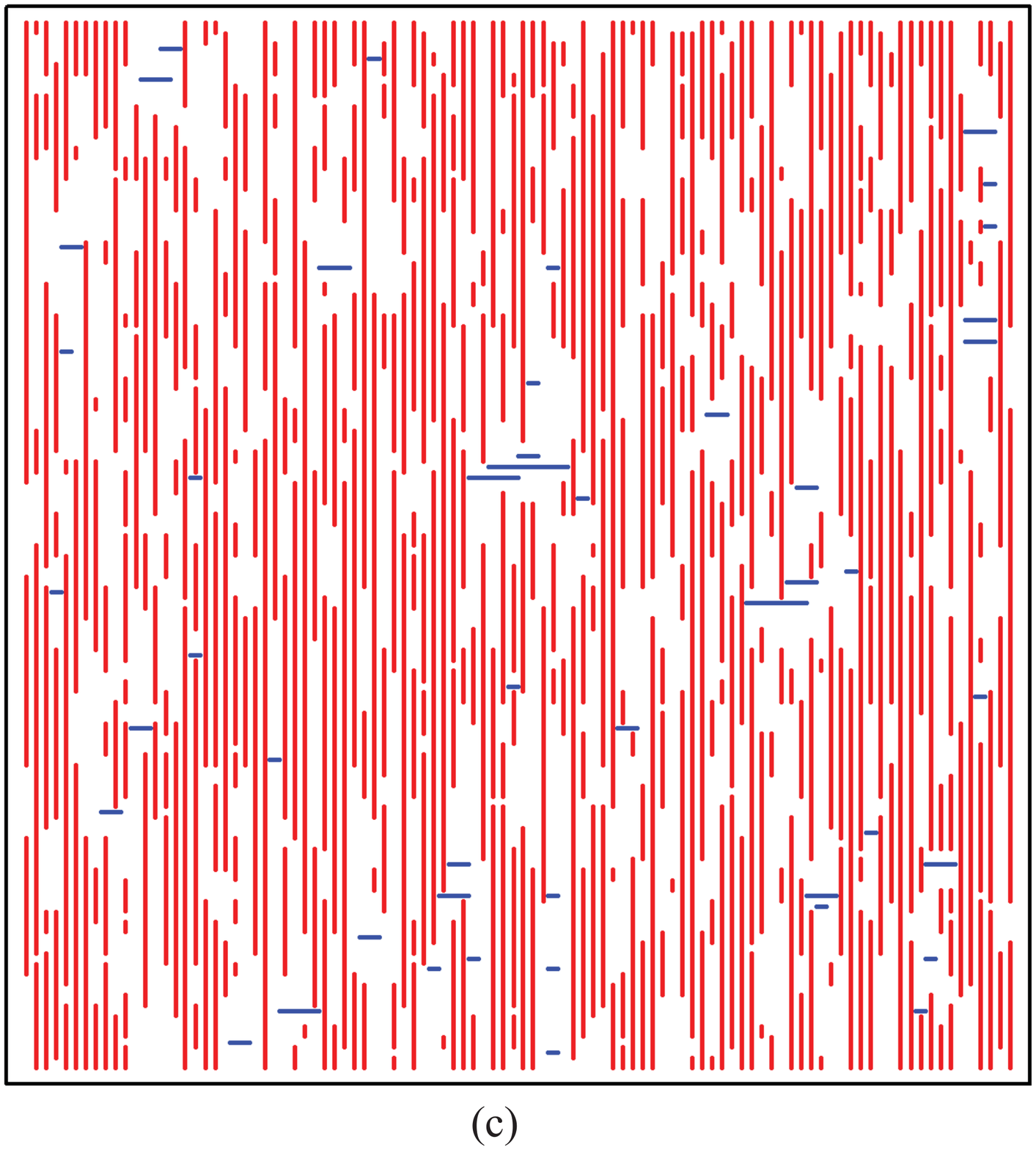}
\caption{} \label{figure3}
\end{figure*}

\newpage

\begin{figure*}[t]
\includegraphics[width=16cm,clip=true]{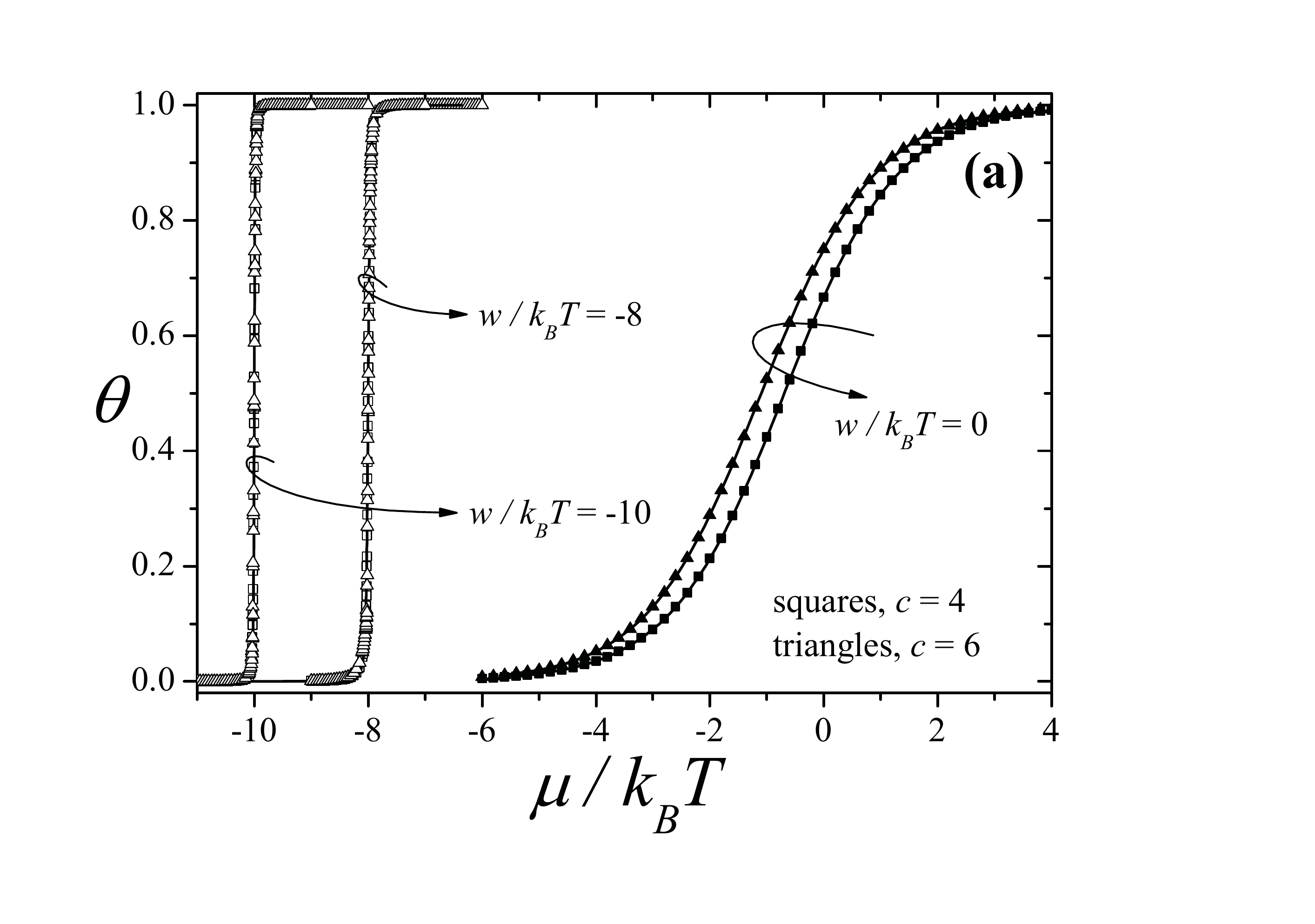}
\includegraphics[width=16cm,clip=true]{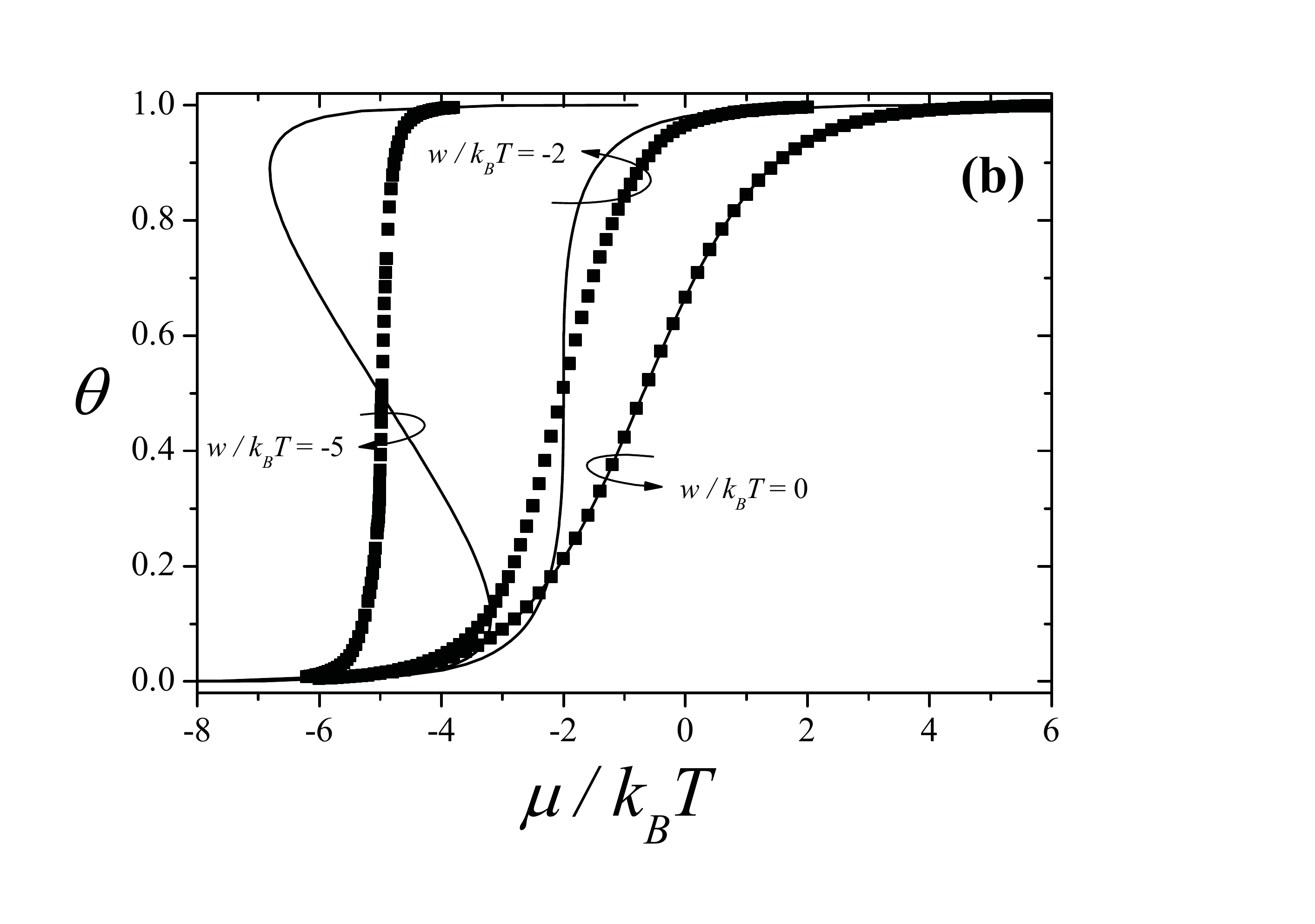}
\caption{} \label{figure4}
\end{figure*}

\newpage

\begin{figure*}[t]
\includegraphics[width=16cm,clip=true]{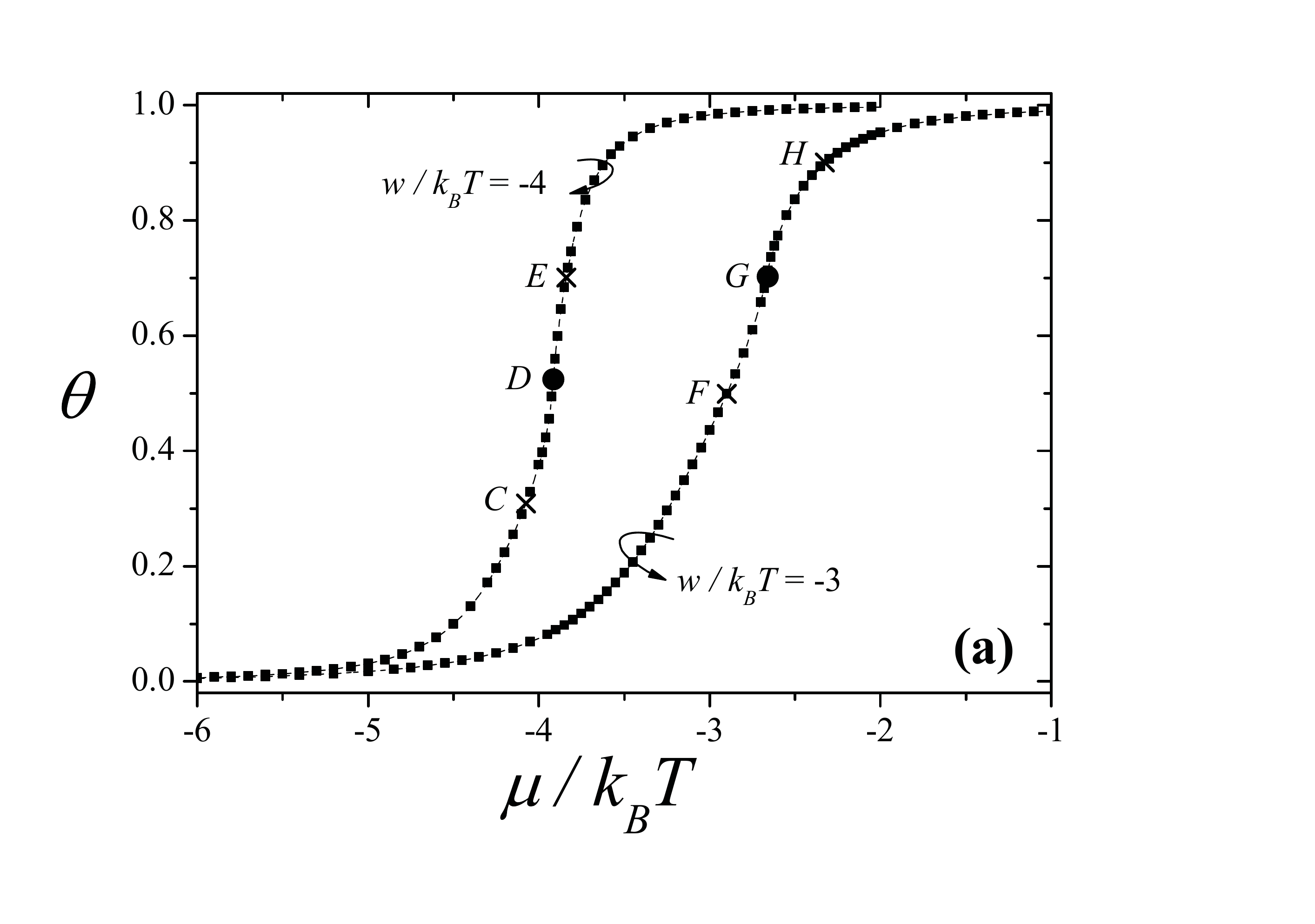}
\includegraphics[width=16cm,clip=true]{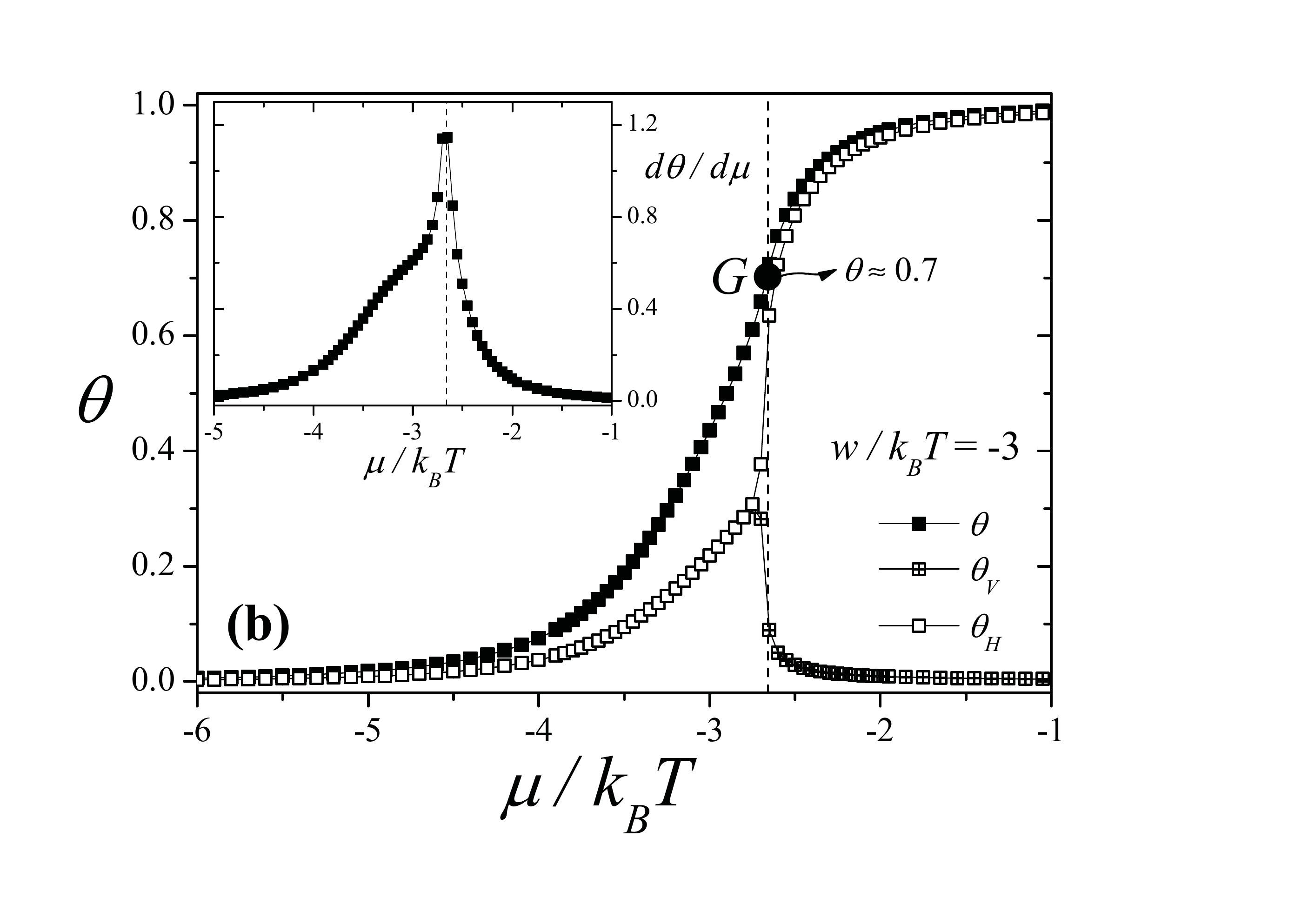}
\caption{} \label{figure5}
\end{figure*}

\newpage

\begin{figure}[t]
\includegraphics[width=16cm,clip=true]{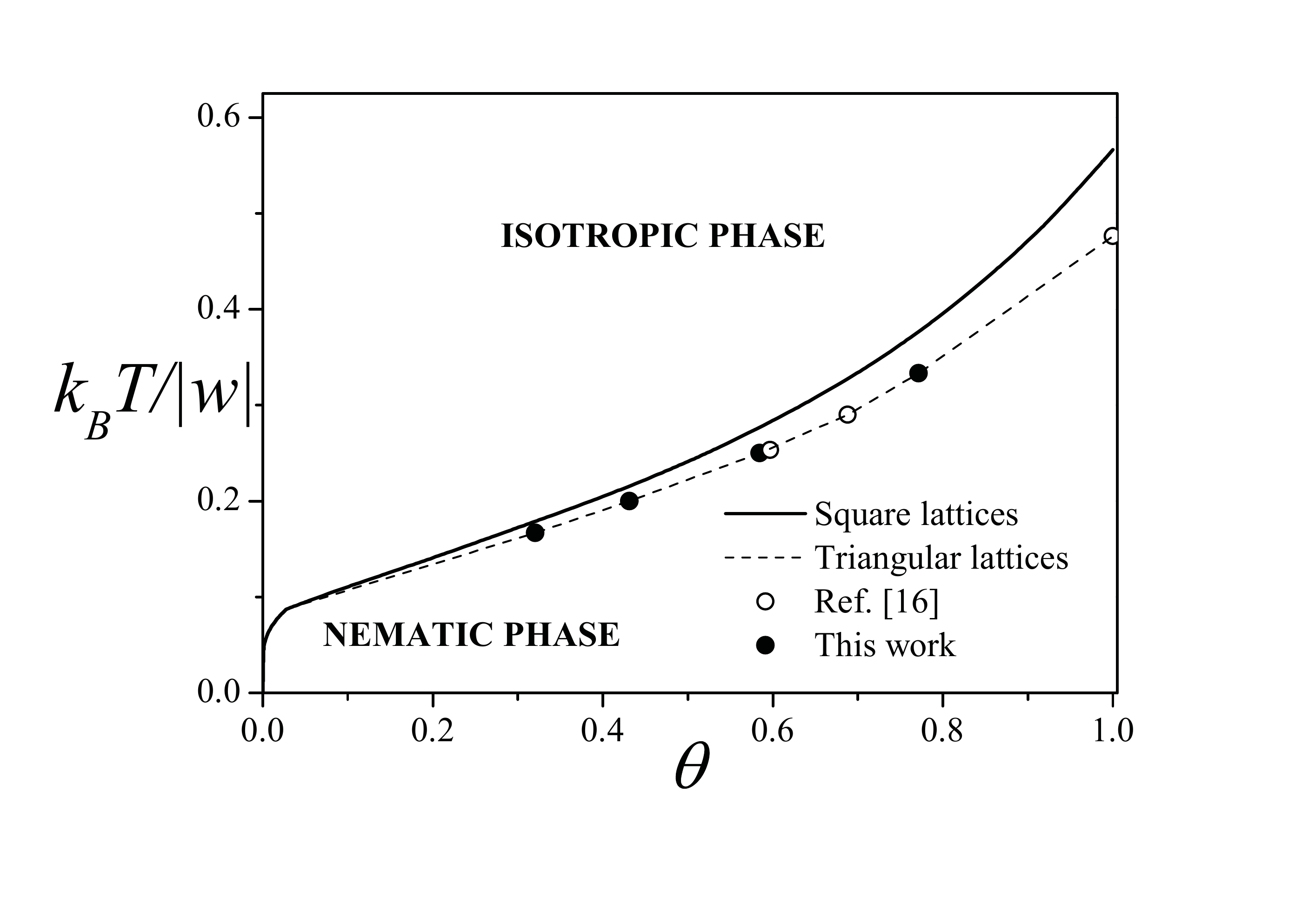}
\caption{}\label{figure6}
\end{figure}


\begin{thebibliography}{99}
\bibitem{foot0} If the operation ${\rm div}$ 1 is not used in \ref{Ham},
there will be three possible values of energy between each pair of
nearest-neighbor monomers on the triangular lattice: $w$ if they
are aligned in the direction of the intermolecular vector
$\vec{r}_{ij}$; $w/2$ if one is aligned in the direction of
$\vec{r}_{ij}$ and the other not; and $w/4$ if they are oriented
in one direction but different to $\vec{r}_{ij}$.
\end{thebibliography}
\end{document}